\documentclass[12pt]{article}
\usepackage{a4wide}
\usepackage{setspace}
\makeatletter

\newcommand{\LyX}{L\kern-.1667em\lower.25em\hbox{Y}\kern-.125emX\@}

\makeatother

\begin{document}

{\par\centering \textbf{\large On the origin of noisy states whose teleportation
fidelity can be enhanced through dissipation }\large \par}

{\par\centering Somshubhro Bandyopadhyay\footnote{
som@ee.ucla.edu, dhom@boseinst.ernet.in
} \par}

{\par\centering \emph{\small Department of Physics, Bose Institute, 93/1 A. P.
C Road, Calcutta - 700009, India}\small \par}

\begin{abstract}
Recently Badziag \emph{et al.} \cite{badziag} obtained a class of noisy states
whose teleportation fidelity can be enhanced by subjecting one of the qubits
to dissipative interaction with the environment via amplitude damping channel
(ADC). We show that such noisy states result while sharing the states \( \left| \Phi ^{\pm }\right\rangle =\frac{1}{\sqrt{2}}\left( \left| 00\right\rangle \pm \left| 11\right\rangle \right)  \)
across ADC. We also show that under similar dissipative interactions different
Bell states give rise to noisy entangled states that are qualitatively very
different from each other in the sense, only the noisy entangled states constructed
from the Bell states  \( \left| \Phi ^{\pm }\right\rangle  \) can \emph{}be
made better sometimes by subjecting the unaffected qubit to a dissipative interaction
with the environment. Importantly if the noisy state is non teleporting then
it can always be made teleporting with this prescription. We derive the most
general restrictions on improvement of such noisy states assuming that the damping
parameters being different for both the qubits. However this curious prescription
does not work for the noisy entangled states generated from \( \left| \Psi ^{\pm }\right\rangle =\frac{1}{\sqrt{2}}\left( \left| 01\right\rangle \pm \left| 10\right\rangle \right)  \).
This shows that an apriori knowledge of the noisy channel might be helpful to
decide which Bell state needs to be shared between Alice and Bob. \emph{}
\end{abstract}

\section{Introduction }

Quantum entanglement \cite{qent}, is a property of bipartite systems by virtue
of which the spatially separated subsystems exhibit strong correlations among
themselves that cannot be explained in classical terms. Often termed as quantum
nonlocality, the subject had been under extensive study and debate since 1935
when Einstein, Podolsky and Rosen questioned the completeness of quantum mechanics
\cite{epr}. Later, in 1964 in his pioneering work John Bell \cite{bell} showed
that such nonlocal correlations are \emph{inherently} quantum mechanical and
any model admitting the description of local hidden variables (or one may prefer
to say local realism) fails to explain such correlations. A consequence of quantum
superposition principle, thus entanglement has been the most celebrated manifestation
of quantum nonlocality \cite{nonlocality}. 

In recent years there is a renewed interest in entanglement primarily because
of its newly found applications in quantum information processing \cite{bennett}
and quantum computing \cite{preskill} and as it turned out, many of the applications
would have either been impossible or less efficient classically. A necessary
condition for faithful implementation of a quantum information protocol is that
the parties share a maximally entangled state (MES), each of them having access
to their respective subsystem whereby they perform local quantum operations
and communicate among themselves one way or both ways via classical channel
to implement the concerned protocol. The local operations include unitary operations,
von neuman measurements and generalized measurements which may involve ancillary
systems Alice and Bob might prepare in their laboratories. For two qubit systems
a class of maximally entangled states are called Bell states that are defined
by
\begin{eqnarray}
\left| \Phi ^{\pm }\right\rangle  & = & \frac{1}{\sqrt{2}}\left( \left| 00\right\rangle \pm \left| 11\right\rangle \right) \label{1} \\
\left| \Psi ^{\pm }\right\rangle  & = & \frac{1}{\sqrt{2}}\left( \left| 01\right\rangle \pm \left| 10\right\rangle \right) \label{2} 
\end{eqnarray}
 The four Bell states are all equally good for faithful quantum communication.
For instance, quantum teleportation \cite{bbcjpw} with fidelity one is achieved
only with maximally entangled states, viz. by any one of the four Bell states.
They also share the property of being local unitary equivalent (they are mapped
onto each other by local pauli rotations). 

As noted above maximally entangled states are crucial for faithful information
processing. In practice, however, entanglement is susceptible to local interactions
with the environment. Such dissipative interactions may take place during encoding/decoding
processes, transmission and sharing of entanglement. This give rise to a mixed
entangled state or a separable state depending on the nature, strength and duration
of the interaction. The noisy states thus formed are of little or no use for
information processing. However one can still generate a fewer number of maximally
entangled states from an ensemble of mixed entangled states by applying the
distillation protocols \cite{distill}. 

Despite being responsible for destroying quantum coherence there is a genuine
positive side of dissipative effects as shown recently by Badziag \emph{et al}.
\cite{badziag}. They demonstrated that a family of non teleporting mixed entangled
states can be made teleporting through dissipative intaeractions with the environment
via amplitude damping channel. Here it may be worth mentioning that ADC is fairly
rich in structure and was shown to be having many interesting properties. For
instance Bennett \emph{et al.} showed that if the noisy channel is ADC then
by entangling the transmission bits one can increase the receiver's capability
of correct inference \cite{fuchs}. 

In the present work we uncover some curious features when a pure but maximal
entanglement interacts with the environment via ADC. The problem that we consider
in this paper is the following. Suppose we have a bipartite maximally entagled
state, say, \( \rho  \) (\( \rho  \) being one of the four Bell states), and
the qubits of the entangled pair undergo local interactions with their respective
environment via amplitude damping channel. There are two possibilities:

\emph{Case 1}: Only one qubit gets affected: \( \rho \rightarrow \rho _{1} \)
(a mixed entangled state).

\emph{Case 2:} Both the qubits are affected: \( \rho \rightarrow \rho _{2} \)
(a mixed entangled state).

Typically, both the cases may arise while sending entanglement across a noisy
channel-the so called transmission errors or due to some coding / decoding process
via some noisy channel. Whatever be the source of errors we ask which of the
two states \( \rho _{1} \) and \( \rho _{2} \), is more useful for teleportation
? 

Since both qubits are affected in the second case, one may be tempted to speculate
that \( \rho _{2} \) is always qualitatively worse that \( \rho _{1} \) in
terms of efficiency to perform teleportation. To our surprise we find that the
state \( \rho _{2} \) is \emph{sometimes} ``better'' than \( \rho _{1} \)
having a higher teleportaion fidelity ! Quite interestingly this effect is observed
\emph{only} if the Bell state that we wish to share is either of the two states
\( \left| \Phi ^{\pm }\right\rangle  \). The effect is also shown to depend
on the strength of the amplitude damping channel. Moreover we show that if \( \rho _{1} \)
is a non teleporting state then it can \emph{always} be made teleporting by
subjecting the other qubit (which didn't interact with the environment before)
to similar dissipative interaction. An interesting feature of this effect is
that the teleportation fidelity of the noisy states that are formed while sharing
the other two Bell states \emph{\( \left| \Psi ^{\pm }\right\rangle  \), cannot}
be enhanced by the same prescription. Here we note that since the process is
trace preserving, we have thus been able to identify and parametrize the class
of non teleporting states \emph{}having fully entangled fraction \( f \) \cite{fef}
less than 1/2, \emph{}for which we can increase \( f \) to greater than half
so that they become suitable for distillation \cite{distill} \emph{without
discarding any pair,} as opposed to filtering procedures \cite{filter1, filter2}.
This shows that \emph{an a priori knowledge of the noisy channel might be helpful
to decide which Bell state needs to be shared between Alice and Bob.} 

As noted earlier, Badziag \emph{et al.} \cite{badziag} obtained a class of
noisy states whose teleportation fidelity can be enhanced by subjecting one
qubit to undergo dissipative interactions with its local environment via ADC.
\emph{What is the origin of such noisy states ?} Our results show that the class
of states obtained in Ref. \cite{badziag} belong to the family of noisy states
that result when any of the particular Bell states, namely \( \left| \Phi ^{\pm }\right\rangle  \)
interacts with the environment via the ADC. 

This paper is arranged in the following way. In Sec. II we review some known
results relating fully entangled fraction to teleportation fidelity and distillation.
In Sec. III the action of an ADC on a qubit is reviewed. The generation of noisy
states while sharing a Bell state across ADC is described in Sec. IV. In Sec.
V we discuss the improvement of noisy states via dissipation. We conclude with
some remarks and discussion in Sec. VI.

\section{Entanglement Fidelity, Fully Entangled Fraction, Teleportation Fidelity and
Distillation threshold }

In this paper reliability for teleportation will be the criterion for judging
the quality the noisy entangled states. The following results will be useful
. 

(1) Fully entangled fraction \( f \) of a bipartite entangled state \( \rho  \)
in \( C^{2}\otimes C^{2} \) is defined as
\begin{equation}
\label{3}
f\left( \rho \right) =\max _{\psi }\left\langle \psi \right| \rho \left| \psi \right\rangle 
\end{equation}
 where the maximum is taken over all maximally entangled states \( \left| \psi \right\rangle  \)
\cite{fef}. For \( \rho  \) to be useful for quantum teleportation we must
have \( f>\frac{1}{2} \) \cite{fef1, fef2}. It was also shown that in the
standard teleportation scheme \cite{bbcjpw} the maximal fidelity \( F \) achievable
is related to \( f \) by \cite{fef2} 
\begin{equation}
\label{4}
F=\frac{2f+1}{3}
\end{equation}
 (2) For the states with \( f\leq \frac{1}{2} \) one cannot directly apply
the distillation protocol by Bennett \emph{et al.} \cite{distill}. For those
states one first resorts to filtering procedures \cite{filter1,filter2} to
enhance the value of \( f \) from \( f\leq \frac{1}{2} \) to \( f>\frac{1}{2} \).

\section{Action of the Amplitude Damping Channel on a Qubit}

Here we briefly review the action of ADC on a qubit. Details can be found in
Ref. \cite{adc}. The amplitude damping channel describes the interaction of
a two level atom with the electromagnetic field (environment). Specifically
the decay of an excited state of a two level atom by spontaneous emission of
a photon in presence of an e.m. field is what modelled by this channel. The
unitary transformation that governs the evolution of the system and the environment
(the environment can be always taken to be in some pure state without any loss
of generality) is defined by,
\begin{eqnarray}
\left| 0\right\rangle _{A}\left| 0\right\rangle _{E} & \rightarrow  & \left| 0\right\rangle _{A}\left| 0\right\rangle _{E}\label{5} \\
\left| 1\right\rangle _{A}\left| 0\right\rangle _{E} & \rightarrow  & \sqrt{1-p}\left| 1\right\rangle _{A}\left| 0\right\rangle _{E}+\sqrt{p}\left| 0\right\rangle _{A}\left| 1\right\rangle _{E}\label{6} 
\end{eqnarray}
 Physically this implies that if an atom is in an excited state \( \left| 1\right\rangle _{A} \),
with probability \( p \) it makes a transition to the ground state \( \left| 0\right\rangle _{A} \)
with the emission of a photon. The environment as a result also makes a transition
from the ``no-photon'' state \( \left| 0\right\rangle _{E} \) to the ``one-photon''
state \( \left| 1\right\rangle _{E} \). Tracing out the environment we obtain
the Kraus operators, with the Kraus operators \( K_{i} \):
\begin{equation}
\label{7}
K_{1}=\left( \begin{array}{cc}
1 & 0\\
0 & \sqrt{1-p}
\end{array}\right) ,K_{2}=\left( \begin{array}{cc}
0 & \sqrt{p}\\
0 & 0
\end{array}\right) 
\end{equation}
 satisfying the completeness relation
\begin{equation}
\label{8}
\sum ^{2}_{i=1}K^{\dagger }_{i}K_{i}=1
\end{equation}
 The density matrix \( \rho  \) of the quantum system then evolves as
\begin{equation}
\label{9}
\rho \longrightarrow \rho ^{\prime }=S\left( \rho \right) =\sum ^{2}_{i=1}K_{i}\rho K^{\dagger }_{i}
\end{equation}
 The above equation basically defines a linear map which takes a density matrix
to another density matrix (a superoperator). It is clear that the amplitude
damping channel is characterized by the parameter \( p \), which denotes the
dissipation strength when a qubit interacts with the environment via this channel.

\section{Sending Bell States across the Amplitude Damping Channel}

In this section we consider the problem of sending the Bell states across the
amplitude damping channel. We assume that Alice prepares one of the Bell states
locally and send one of the qubits to Bob. This qubit interacts with the environment
via ADC during the transmission. The initial Bell state is thus transformed
to a mixed entangled state. Let \( \rho  \) be the density operator representing
one of the four Bell states, then the interaction with the environment via ADC
is described by the following transformation:
\begin{equation}
\label{10}
\rho \longrightarrow \widetilde{\rho }\left( p\right) =S\left( \rho \right) =W_{0}\rho W^{\dagger }_{0}+W_{1}\rho W^{\dagger }_{1}
\end{equation}
 where \( W_{i}\equiv I\otimes K_{i} \) are given by
\begin{equation}
\label{11}
W_{0}=\left[ \begin{array}{cccc}
1 & 0 & 0 & 0\\
0 & \sqrt{1-p} & 0 & 0\\
0 & 0 & 1 & 0\\
0 & 0 & 0 & \sqrt{1-p}
\end{array}\right] ;W_{1}=\left[ \begin{array}{cccc}
0 & \sqrt{p} & 0 & 0\\
0 & 0 & 0 & 0\\
0 & 0 & 0 & \sqrt{p}\\
0 & 0 & 0 & 0
\end{array}\right] 
\end{equation}
 The following notation will be used henceforth to represent the Bell states
\begin{equation}
\label{12}
\left| \Phi ^{\pm }\right\rangle \left\langle \Phi ^{\pm }\right| \equiv \rho _{\pm }^{\Vert };\left| \Psi ^{\pm }\right\rangle \left\langle \Psi ^{\pm }\right| \equiv \rho _{\pm }^{\bot }
\end{equation}
 It is clear that such an interaction turns the maximally entangled state, originally
shared by Alice and Bob into a mixed entangled state,
\begin{equation}
\label{13}
\widetilde{\rho }_{\pm }^{\Vert }=\left[ \begin{array}{cccc}
1 & 0 & 0 & \pm \sqrt{1-p}\\
0 & 0 & 0 & 0\\
0 & 0 & p & 0\\
\pm \sqrt{1-p} & 0 & 0 & 1-p
\end{array}\right] :\widetilde{\rho }_{\pm }^{\bot }=\frac{1}{2}\left[ \begin{array}{cccc}
p & 0 & 0 & 0\\
0 & 1-p & \pm \sqrt{1-p} & 0\\
0 & \pm \sqrt{1-p} & 1 & 0\\
0 & 0 & 0 & 0
\end{array}\right] 
\end{equation}
 It turns out that the fully entangled fraction \( f \) of the four resulting
mixed states is the same:
\begin{equation}
\label{14}
f\left( \widetilde{\rho }_{\pm }^{\parallel }\right) =f\left( \widetilde{\rho }_{\pm }^{\perp }\right) =\frac{1}{4}\left( 1+\sqrt{1-p}\right) ^{2}=f\left( \widetilde{\rho }\right) 
\end{equation}
 where
\begin{equation}
\label{15}
f\left( \widetilde{\rho }_{\pm }^{\Vert }\right) =\left\langle \Phi ^{\pm }\right| \widetilde{\rho }_{\pm }^{\Vert }\left| \Phi ^{\pm }\right\rangle ;\quad f\left( \widetilde{\rho }_{\pm }^{\bot }\right) =\left\langle \Psi ^{\pm }\right| \widetilde{\rho }_{\pm }^{\bot }\left| \Psi ^{\pm }\right\rangle 
\end{equation}
 Here we note that fully entangled fraction is same as that of entanglement
fidelity \cite{schumacher}. Of course this is not always the case, as we will
show later when we subject the second qubit to dissipation for possible improvement
of the noisy states. Since entanglement fidelity measures how well the entanglement
has been preserved while interacting with a noisy channel we conclude that till
this end all the four Bell states are equally corrupted. This is also expected
as the noisy channel acts locally on one of the qubits whose state given by
the reduced density matrix is the same for all the four Bell states. 

We now consider the usefulness of the mixed states given by Eq. (15) to perform
quantum teleportation. These family of mixed states are useful for teleportation
when \( f\left( \widetilde{\rho }\right) >1/2 \) \cite{fef1, fef2}, from which
one easily obtains the restriction on the parameter \( p \)
\begin{equation}
\label{16}
\begin{array}{cc}
f\left( \widetilde{\rho }\right) >\frac{1}{2} & \forall p<2\sqrt{2}-2
\end{array}
\end{equation}
 Eq. (16) simply states that for all values of \( p\geq 2\sqrt{2}-2 \), \( f\left( \widetilde{\rho }\right) \leq 1/2 \)
the family of mixed states \( \widetilde{\rho }_{\pm }^{\parallel },\, \widetilde{\rho }_{\pm }^{\perp } \)
so obtained are not useful for teleportation. Here ``not useful'' is understood
as not better than what can be done classically.

\section{Improving the Noisy Bell States by letting the unaffected Qubit to interact
with the Local Environment via Amplitude Damping Channel}

In this section we discuss the possible improvement of the noisy states given
by Eq. (15) by subjecting the second qubit to similar dissipative interaction.

\subsection{Damping Parameters same for the two qubits}

We now allow Alice's qubit to interact with the environment via the amplitude
damping channel. Here we also assume that the strength of dissipation affecting
qubits locally is the same for both. This is not a necessary assumption but
only simplifies the degree of algebraic complexity though it does not capture
all the intricacies involved. The general case where the damping parameter is
different for the two qubits will be treated in the next subsection. 

This interaction is described by the following transformation:
\begin{equation}
\label{17}
\widetilde{\rho }(p)\longrightarrow \widetilde{\widetilde{\rho }}(p)=S^{\prime }\left( \widetilde{\rho }\right) =W^{\prime }_{0}\rho W^{\prime \dagger }_{0}+W^{\prime }_{1}\rho W^{\prime \dagger }_{1}
\end{equation}
 where \( W^{\prime }_{i}=K_{i}\otimes I \) and \( \widetilde{\rho } \) is
the density operator representing one of the noisy Bell states: \( \left\{ \widetilde{\rho }_{\pm }^{\parallel },\widetilde{\rho }_{\pm }^{\bot }\right\}  \). 

The fully entangled fraction of the \( \widetilde{\widetilde{\rho }} \) states
are now given by:
\begin{eqnarray}
f\left( \widetilde{\widetilde{\rho }}_{\pm }^{\parallel }\right)  & = & 1-p+\frac{1}{2}p^{2}\label{18} \\
f\left( \widetilde{\widetilde{\rho }}_{\pm }^{\bot }\right)  & = & \begin{array}{cc}
1-p & \forall p\leq \frac{2}{3}\\
\frac{p}{2} & \forall p\geq \frac{2}{3}
\end{array}\label{19} 
\end{eqnarray}
 First we check if there is a range of \( p \) such that \( f\left( \widetilde{\widetilde{\rho }}\right) >1/2 \).
If so then we would like to know if that range has an overlap with the range
\( p\geq 2\sqrt{2}-2 \) because only then that would imply that further application
of a dissipative interaction can indeed enhance the teleportation fidelity of
the ``one qubit affected'' noisy states that are formed via the interaction
defined by Eq. (10). It is easy to check that 
\begin{eqnarray}
f\left( \widetilde{\widetilde{\rho }}_{\pm }^{\parallel }\right)  & >\frac{1}{2} & \forall p,p\neq 1\textrm{ }\label{20} \\
\textrm{ }f\left( \widetilde{\widetilde{\rho }}_{\pm }^{\bot }\right) = & \begin{array}{cc}
1-p & >\frac{1}{2}\\
\frac{p}{2} & <\frac{1}{2}
\end{array} & \begin{array}{cc}
\forall p<\frac{1}{2} & \\
\forall p,p\neq 1
\end{array}\label{21} 
\end{eqnarray}
 Recall that for all \( p\geq 2\sqrt{2}-2 \), \( f\left( \widetilde{\rho }\right) \leq 1/2 \).
Since from Eq. (20) it follows that \( f\left( \widetilde{\widetilde{\rho }}_{\pm }^{\parallel }\right) >\frac{1}{2}\; \forall p,p\neq 1\textrm{ } \),
therefore \( f\left( \widetilde{\widetilde{\rho }}_{\pm }^{\parallel }\right) \textrm{ }>\frac{1}{2}\quad \forall p\geq 2\sqrt{2}-2 \)
which implies that teleportation fidelity of \emph{all} non teleporting states
\( \widetilde{\rho }_{\pm }^{\parallel }\left( p\right) ;\; p\geq 2\sqrt{2}-2 \)
can indeed be enhanced by subjecting the second qubit to dissipation. Here we
also note that \( f\left( \widetilde{\widetilde{\rho }}_{\pm }^{\parallel }\right) \geq f\left( \widetilde{\rho }_{\pm }^{\parallel }\right) \; \forall p\geq 0.80585 \).
This implies that teleportation fidelity of the teleporting states \( \widetilde{\rho }_{\pm }^{\parallel }\left( p\right) ;\; 0.80585\leq p<2\sqrt{2}-2 \)
can also be enhanced by this method. Thus the two-qubit affected state \( \widetilde{\widetilde{\rho }}_{\pm }^{\Vert } \)
is \emph{better} (i.e. improved) with a higher teleportation fidelity than single
qubit affected state \( \widetilde{\rho }_{\pm }^{\Vert } \) whenever \( p>0.80585 \).
This is surprising in the sense that the dissipative interaction which spoils
entanglement in the first step is utilised to improve the quality of the mixed
state by applying it to the second subsystem. 

However Eqs. (19) and (21) show that the same prescription though curious doesn't
work to turn non teleporting states into teleporting ones. Although \( f\left( \widetilde{\widetilde{\rho }}_{\pm }^{\bot }\right) =p/2\geq f\left( \widetilde{\rho }_{\pm }^{\bot }\right) \; \forall p\geq 8/9 \),
the enhancement is not sufficient to transform non teleporting states to teleporting
states as \( f\left( \widetilde{\widetilde{\rho }}_{\pm }^{\bot }\right) =p/2 \)
is always less than or equal to \( \frac{1}{2} \). 

It may be worth stressing that in showing the above effect we have taken the
damping parameter to be the same for both the qubits. We will now consider the
case where the damping parameters are different.

\subsection{Damping parameters different for the two qubits}

Let us now treat the problem in a more general way. Here, as before, we will
assume that first Bob's qubit undergoes dissipative interaction with the environment
and the parameter \emph{p} of the amplitude damping channel will now be denoted
by \( p_{b} \). After that we allow the qubit that belongs to Alice also to
interact with the environment via amplitude damping channel and the channel
parameter now taken to be different from \( p_{b} \), is denoted by \( p_{a} \).
Nevertheless the conclusion remains the same except for the fact that in this
case we have the freedom of varying the strength of dissipation for the second
qubit which allows us to maximize the enhancement of teleportation fidelity. 

The fully entangled fraction \( f\left( \widetilde{\rho }\right)  \) of the
resulting mixed state \( \widetilde{\rho }(p_{b}) \) formed via interaction
given by Eq. (10) is,
\begin{equation}
\label{22}
f\left( \widetilde{\rho }_{\pm }^{\parallel }\right) =f\left( \widetilde{\rho }_{\pm }^{\perp }\right) =\frac{1}{4}\left( 1+\sqrt{1-p_{b}}\right) ^{2}=f\left( \widetilde{\rho }\right) 
\end{equation}
 This is same as Eq. (14), the only difference is that \( p \) is now denoted
by \( p_{b} \). Now when we allow Alice's qubit also to interact with the environment
described by Eq. (17) , \( \widetilde{\rho }(p_{b}) \) states are transformed
to \( \widetilde{\widetilde{\rho }}\left( p_{a},p_{b}\right)  \). 

Irrespective of the source state the fully entangled fraction is the same when
only one qubit is affected. The remaining analysis involving subjection of the
second qubit to dissipation for possible enhancement of teleportation fidelity
is divided into two parts. The first one corresponds to the cases when the source
states are \( \left| \Phi ^{\pm }\right\rangle \left\langle \Phi ^{\pm }\right| \equiv \rho _{\pm }^{\Vert } \)
and the second part corresponds to the cases when the source states are \( \left| \Psi ^{\pm }\right\rangle \left\langle \Psi ^{\pm }\right| \equiv \rho _{\pm }^{\bot } \).

\subsubsection{Part 1}

Let us first recall that for \( p_{b}\geq 2\sqrt{2}-2 \), \( f\left( \widetilde{\rho }\right) \leq 1/2 \),
i.e., the states \( \widetilde{\rho }_{\pm }^{\parallel } \) are non teleporting.
After the second interaction for which the damping parameter is \( p_{a} \),
the fully entangled fraction corresponding to the \( \widetilde{\widetilde{\rho }}^{\parallel }_{\pm }\left( p_{a},p_{b}\right)  \)
states are given by:
\begin{equation}
\label{23}
f\left( \widetilde{\widetilde{\rho }}_{\pm }^{\parallel }\right) =\frac{1}{4}\left[ p_{a}p_{b}+\left( 1+\sqrt{\left( 1-p_{a}\right) \left( 1-p_{b}\right) }\right) ^{2}\right] 
\end{equation}
\textbf{5.2.1.1 Condition for ``two qubit affected'' state having larger teleportation
fidelity than the ``one qubit affected'' state}

Our objective is to obtain the condition such that the inequality
\begin{equation}
\label{24}
f\left( \rho ^{\prime \prime }\right) >f\left( \rho ^{\prime }\right) 
\end{equation}
 is satisfied. To be precise we are now looking for those values of \( p_{a} \)
, such that for \( p_{b}\geq 2\sqrt{2}-2 \), the previous inequality is satisfied.
 It turns out that for
\begin{equation}
\label{25}
p_{a}<\frac{4\left[ \sqrt{1-p_{b}}\left( 2p_{b}-1\right) -\left( 1-p_{b}\right) \right] }{\left( 2p_{b}-1\right) ^{2}}=g\left( p_{b}\right) 
\end{equation}
 the inequality \( f\left( \rho ^{\prime \prime }\right) >f\left( \rho ^{\prime }\right)  \)
is satisfied. However it remains to be ensured that a range of values of \( p_{a} \)
is obtained for \( p_{b}\geq 2\sqrt{2}-2 \). First observe that \( g(3/4)=0 \).
This in turn implies we can always find some suitable range of \( p_{a} \)
such that the inequality (20) is satisfied provided \( p_{b}>3/4 \). Stated
more explicitly this means that only when Bob's qubit interacted with the environment
via the amplitude damping channel characterized by the parameter \( p_{b}>3/4 \)
, then the possibility of improving the corrupted state by allowing the other
qubit to interact, arises. Hence the two qubit affected states are thereby made
better than the one qubit affected state. One may also note that not only non
teleporting states can be made teleporting but also teleporting states with
poor fidelity are made better. We now look for the condition when maximum enhancement
of teleportation fidelity is achieved. 

\textbf{5.2.1.2 Maximum achievable teleportation fidelity}

Let us now assume \( p_{b}>3/4 \). Thus we have at our disposal a range of
possible values of \( p_{a} \) satisfying (24) and the inequality (25). Then
another important question remains to be answered: For which value of \( p_{a} \),
\( f\left( \rho ^{\prime \prime }\right)  \) is the maximum? That is, for which
value \( p_{a} \) highest fidelity for teleportation can be achievd?  

One can easily check that (23) is maximised for
\begin{equation}
\label{26}
p_{a}=\frac{p_{b}\left( 4p_{b}-3\right) }{\left( 2p_{b}-1\right) ^{2}}
\end{equation}
 Here it is interesting to note that (26) immediately gives a lower bound  on
the value of the parameter \( p_{b} \), which is 3/4 and this bound we have
obtained before. Substituting (26) in (23) one obtains the maximum value of
fully entangled fraction that can be achieved:
\begin{equation}
\label{27}
f_{max}\left( \rho ^{\prime \prime }\right) =\frac{p^{2}_{b}}{2\left( 2p_{b}-1\right) }
\end{equation}

One should note that \( f_{max}>1/2 \). 

\textbf{5.2.1.3 Non teleporting states can always be made teleporting }

To begin with Alice and Bob initially shared a maximally entangled state. If
Bob's qubit gets affected by the ADC , \( p_{b}>3/4 \) then the fidelity of
the resulting noisy state can be improved by subjecting the unaffected qubit
to similar dissipation characterized by the parameter \( p_{a} \)  in accordance
with Eq. (25). The maximum ``fully entangled fraction'' or the maximum fidelity
of teleportation is obtained by substituting (26) in the expression (23). 

Since for \( p_{b}\geq 2\sqrt{2}-1 \) the state \( \rho ^{\prime }(p_{b}) \)
is not capable of teleportation better than what can be achieved classically.
It is now clear that those non teleporting states can always be made teleporting
by allowing the other qubit to interact with the environment with an appropriate
choice of the parameter \( p_{a} \). In particular the maximum fidelity can
be obtained if we choose \( p_{a} \) in accordance with (26). Even if the state
\( \rho ^{\prime }(p_{b}) \) is a teleporting one (which it is, if \( 3/4<p_{b}<2\sqrt{2}-2 \))
still its fidelity can be enhanced by allowing the other qubit to interact with
the environment for any value of \( p_{a} \) defined by (25).

\subsubsection{Part 2}

We now analyze what happens when the source states are the superposition of
antiparallel spins. First we note that when one qubit is affected the fully
entangled fraction is given by Eq. (10) or Eq. (22). After the second qubit
is affected, the interaction being governed by an appropriate superoperator,
the density operator of the mixed state is given by 
\[
\widetilde{\widetilde{\rho }}_{\pm }^{\bot }=\left[ \begin{array}{cccc}
p_{a}+p_{b} & 0 & 0 & 0\\
0 & 1-p_{b} & \pm \sqrt{\left( 1-p_{a}\right) \left( 1-p_{b}\right) } & 0\\
0 & \pm \sqrt{\left( 1-p_{a}\right) \left( 1-p_{b}\right) } & 1-p_{a} & 0\\
0 & 0 & 0 & 0
\end{array}\right] \]

Now recall the definition of fully entangled fraction. One can easily see that
two possible candidates for fully entangled fraction are 

\begin{eqnarray*}
f_{1} & = & \frac{p_{a}+p_{b}}{4}\\
f_{2} & = & \frac{1}{4}\left( \sqrt{1-p_{a}}+\sqrt{1-p_{b}}\right) ^{2}
\end{eqnarray*}

The fully entangled fraction is given by \( \max \left[ f_{1},f_{2}\right]  \).
Whether \( f_{1}>f_{2} \) or vice versa depends on the particular values of
\( p_{a} \) and \( p_{b} \), i.e., whether \( f_{1} \) is the fully entangled
fraction or \( f_{2} \) is depends on the specific choices of the damping parameters. 

Let us first assume that \( f_{1}>f_{2} \). Now observe that \( f_{1}\leq 1/2 \).
Therefore teleportation fidelity can never exceed 2/3. Hence non teleporting
states can never become teleporting. Now if \( f_{2}>f_{1} \), then note that
\( f_{2}(p_{a},p_{b})\leq f\left( \widetilde{\rho }_{\pm }^{\perp }\right) \; \forall p_{a},p_{b} \).
Therefore whatever be the case the above two observations guarantee that by
subjecting the second qubit to a similar dissipative interaction, the non teleporting
states cannot be made teleporting.

\section{Remarks and Conclusion}

We know that given any mixed state the corresponding teleportation fidelity
has contribution from two parts: (a) the quantum part i.e. the shared entanglement
and (b) the classical part, i.e. the classical correlations. Since entanglement
cannot be increased by TPLOCC, an enhancement  in teleportation fidelity suggests
that, for the states on the border line of non teleporting and teleporting,
it is the classical part that contributes more heavily than the quantum part
so that even a non teleporting state becomes teleporting in spite of losing
entanglement. Thus TPLOCC which spoils entanglement but nevertheless can sometimes
redistribute the classical correlations in a helpful way. 

We now discuss the implications of our result in the context of distilling entanglement
from the noisy states. For distillation \cite{distill} of mixed entangled states
having fully entangled fraction \( f\leq \frac{1}{2} \)one needs to apply filtering
methods \cite{filter1,filter2} so that \( f \) exceeds \( \frac{1}{2} \).
This is important because only then the distillation protocol of Bennett et
al \cite{distill} could be applied. However filtering being non trace preserving
involves measurement followed by post selection which necessarily discards a
large number of available pairs. In view of this our results provide a significant
improvement for a large class of states for which fully entangled fraction can
be made greater than half without resorting to filtering implying that we need
not discard pairs. 

We have shown that the noisy entangled states constructed from the source states
by similar (meaning same for all the source states) dissipative interactions,
can indeed have very different properties even though the source states are
local unitary equivalent. This observation might be helpful while trying to
establish entanglement when the noisy quantum channel is the amplitude damping
channel. If we are provided with either of the states \( \left| \Psi ^{\pm }\right\rangle  \)
then instead of trying to share it we may first convert it to \( \left| \Phi ^{\pm }\right\rangle  \)
state by applying pauli rotations locally. The reason is even if the entanglement
gets spoiled at its worse so that the mixed states are non teleporting we can
still make them suitable for teleportation by TPLOCC. Besides being made useful
for teleportation these states also allow for direct application of the Bennett
protocol for distillation \cite{distill}. Thus our results do indicate that
a priori knowledge of the noisy channel may decide sometimes which Bell state
is to be shared. One of course should try to share the case specific Bell state
for which the resulting mixed states allow for further useful manipulations. 

I would like to thank Daniel Terno, Ujjwal Sen and Guruprasad Kar for helpful
discussions.

\end{document}